\documentclass[letterpaper, 10 pt, conference]{ieeeconf}  

\IEEEoverridecommandlockouts                              

\usepackage{graphics} 
\usepackage{epsfig} 
\usepackage{mathptmx} 
\usepackage{times} 
\usepackage{amsmath} 
\usepackage{amssymb}  
\usepackage{algorithm}
\usepackage{siunitx}
\usepackage[dvipsnames]{xcolor}
\usepackage{algpseudocode}
\usepackage{array}
\usepackage{cite}

\title{\LARGE \bf
A Two-Stage Bayesian Optimisation for Automatic Tuning of an Unscented Kalman Filter for Vehicle Sideslip Angle Estimation*
}

\author{Alberto Bertipaglia$^{1}$, Barys Shyrokau$^{1}$, Mohsen Alirezaei$^{2}$ and Riender Happee$^{1}$
\thanks{*The Dutch Science Foundation NWO-TTW supports the research within the EVOLVE project (nr. 18484).}
\thanks{$^{1}$Alberto Bertipaglia, Barys Shyrokau and Riender Happee are with the Department of Cognitive     Robotics, Delft University of Technology, 2628 CD Delft, The Netherlands
        {\tt\small A.Bertipaglia@tudelft.nl}, \tt\small {B.Shyrokau@tudelft.nl} and \tt\small {R.Happee@tudelft.nl}}%
\thanks{$^{2}$Mohsen Alirezaei is with the Department of Mechanical Engineering, University of Eindhoven, 5612 AZ Eindhoven, The Netherlands
        {\tt\small m.alirezaei@tue.nl}}%
}

\begin{document}

\maketitle
\thispagestyle{empty}
\pagestyle{empty}

\begin{abstract}

This paper presents a novel methodology to auto-tune an Unscented Kalman Filter (UKF). It involves using a Two-Stage Bayesian Optimisation (TSBO), based on a t-Student Process to optimise the process noise parameters of a UKF for vehicle sideslip angle estimation. Our method minimises performance metrics, given by the average sum of the states' and measurement' estimation error for various vehicle manoeuvres covering a wide range of vehicle behaviour. The predefined cost function is minimised through a TSBO which aims to find a location in the feasible region that maximises the probability of improving the current best solution. Results on an experimental dataset show the capability to tune the UKF in \SI{79.9}{\%} less time than using a genetic algorithm (GA) and the overall capacity to improve the estimation performance in an experimental test dataset of \SI{9.9}{\%} to the current state-of-the-art GA. \looseness = -1

\end{abstract}


\section{INTRODUCTION}

An accurate and real-time estimation of the vehicle sideslip angle is essential to strengthen the performance of active vehicle control systems. The state-of-the-art estimation of sideslip angle mostly relies on model-based approaches using an Unscented Kalman Filter (UKF). Despite its efficiency and robustness, the process and observation noise parameters must be accurately tuned to achieve good performance \cite{chen2018weak, abbeel2005discriminative}. The noise parameters selection is particularly relevant because they need to capture the following aspects:
\begin{itemize}
    \item Process and observation model mismatch with respect to the real vehicle dynamics.
    \item Discretisation error because the time-step in the Kalman filter influences the noise parameters to achieve the best state estimation \cite{chen2021time}.
    \item Different working conditions of the sensors installed in the vehicle.
\end{itemize} 
If the tuning is obtained through numerical optimisation, the cost function is highly non-convex, non-smooth and nonlinear, calling for robust optimisation methods \cite{chen2018weak}.

Several optimisation algorithms for UKF tuning \cite{acosta2019optimized} have been studied: GA, Sequential Quadratic Programming, Nelder-Mead, Artificial Bee Colony, Fruit Fly Optimisation (FFO) and Differential Evolution. The comparison demonstrates that each algorithm leads to different optimum locations, illustrating the importance of the optimization method. However, a comparison regarding the amount of data and the optimisation time has been omitted. Furthermore, only stochastic optimisation algorithms have been compared that can be inefficient if the UKF takes a long simulation time. For vehicle sideslip angle estimation, large-scale training sets representing vehicle behaviour \cite{mazzilli2021benefit} make the cost function evaluation very time-consuming. In different engineering applications, several solutions have been proposed, e.g. Reinforcement Learning (RL) \cite{tang2021reinforcement} or Bayesian Optimisation (BO) \cite{chen2018weak}. However, the majority of these are tested only on simulation data or toy examples. A more detailed analysis of previous works is in Section II. \newline

\begin{figure}[t]
   \centering
   \setlength{\fboxrule}{0pt}
   \framebox{\parbox{3in}{\includegraphics[scale=0.35]{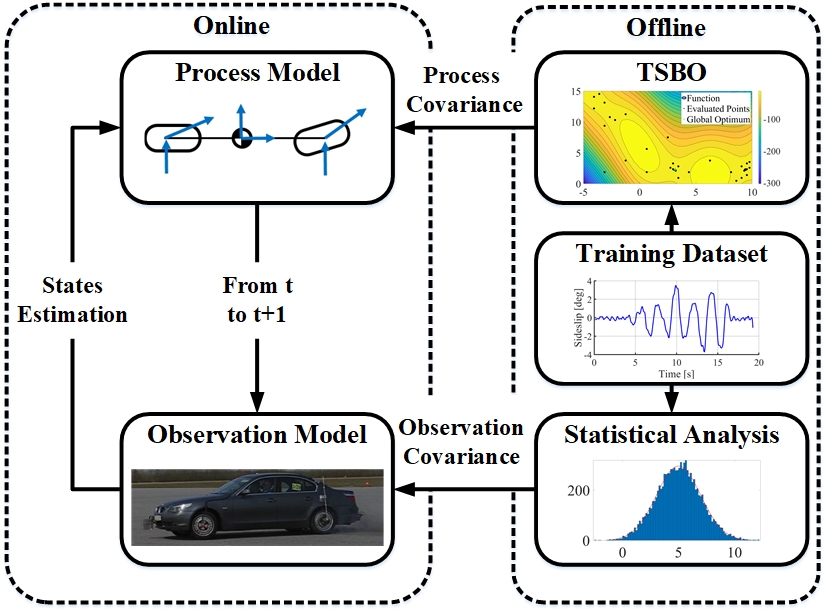}}}
    \caption{Framework utilizing a Two-Stage Bayesian Optimisation (TSBO) to tune the process noise parameters of a UKF.}
    \label{Framework}
\end{figure}

This paper proposes a new UKF tuning methodology using Two-Stage Bayesian Optimisation (TSBO) \cite{torun2018global}, based on a t-Student Process (tSP) for vehicle sideslip angle estimation. The proposed methodology to tune the process noise parameters reduces the optimisation time and improves the optimum localisation. Both process and observation noise are assumed Gaussian, zero mean and uncorrelated, as for the majority of the model-based vehicle state estimation filters \cite{heidfeld2020ukf}. The methodology framework is represented in Figure \ref{Framework}, a more detailed explanation will be presented in Section IV. The observation noise parameters are tuned by performing a statistical analysis of the vehicle sensor measurements \cite{mazzilli2021benefit}. The process noise parameters' tuning is based on optimising a cost function defined as the sum of the Normalized Root Mean Squared Errors (NRMSE) of the vehicle states and measurements for 8 manoeuvres. The 8 manoeuvres which compose the training set and the 23 forming the test set are selected from a real-world experimental dataset composed of 216 manoeuvres recorded at the Automotive Testing Papenburg GmbH. The manoeuvres are chosen to cover a wide range of vehicle motions. The cost function is optimised using the modified TSBO 
\cite{torun2018global}. It consists of a fast exploration and a pure exploitation stage to find the optimum, reducing the number of simulations required. Differently from the TSBO proposed in \cite{torun2018global}, the surrogate model is a tSP to enhance the robustness and the ability to capture heavy-tailed NRMSE. The methodology is experimentally validated using a UKF, but can also be applied to an Extended Kalman Filter (EKF). \looseness = -1

\section{PREVIOUS WORKS}

Despite the extensive use of UKF for vehicle sideslip angle estimation, the tuning of the process and observation noise parameters is rarely investigated in the literature. The observation noise parameters are usually tuned through statistical analysis, obtained after sensor calibration tests \cite{mazzilli2021benefit}. The only exception is when the observation noise parameters are tuned together with the process noise parameters \cite{heidfeld2021optimization}. This rarely happens because the filter performance depends mainly on the ratio between the eigenvalues of the observation and process covariance matrices. Thus, it is more convenient to fix the observation noise parameters, which can be determined through sensor calibration, and optimise the process noise parameters.The methodologies for process noise parameters tuning can be split into three different categories: \emph{manual tuning}, \emph{metaheuristic} optimisation and \emph{data-driven} techniques. An overview of optimisation algorithms for UKF tuning is shown in Table \ref{Table_Overview}. \looseness = -1 

\emph{Manual tuning} is a trial-and-error approach driven by the user experience. The idea is to perform a grid search in the parameter space and reach the best tuning iteratively. Despite the oversimplicity of the concept, it is still very often used, especially for vehicle sideslip angle estimation \cite{van2018adaptive}, because the performance of Kalman filters is stable for a range of process noise parameters settings \cite{mazzilli2021benefit}. 
The main drawback is the inefficiency of the approach.
Moreover, there is no proof to reach the optimal performance. This leads to an unreliable comparison when different filter architectures are analysed.

Thus, Kalman filter tuning with numerical optimisation techniques are widely used. Gradient-based optimisations are the most straightforward algorithms which can be implemented \cite{kerst2019}, but the non-convex and nonlinear cost functions let them easily be trapped into local minima. Thus, gradient-free optimisation algorithms are introduced. One of the simplest algorithms is the coordinate descent algorithm \cite{abbeel2005discriminative} which consists of a cycling increase or decrease of each parameter of the process covariance. 
A similar optimisation technique is the downhill simplex algorithm \cite{powell2002automated}, which consists of evaluating the cost function at the simplex vertices of the predefined sample space. After this, the algorithm should converge to the optimum through a series of movements: reflection, expansion, and contraction. Interested reader can find more information on both algorithms in \cite{abbeel2005discriminative} and \cite{powell2002automated} respectively. These algorithms may easily fall in local optima when the cost function is non-smooth, so more advanced metaheuristic optimisations have been proposed.

\emph{Metaheuristics} are procedures that can provide an acceptable solution to an optimisation problem with incomplete information about the cost function. Examples of metaheuristic algorithms are: the already cited downhill simplex algorithm \cite{powell2002automated}, GA \cite{oshman2000optimal} or the Multi-Objective GA \cite{mazzilli2021benefit}, SA \cite{heidfeld2021optimization} and FFO \cite{acosta2019optimized}. A comparison of different metaheuristic algorithms for Kalman filter tuning \cite{acosta2019optimized} shows that GA and FFO found the best cost function optimum. This explains why GA is the current state-of-the-art for UKF tuning for vehicle sideslip angle estimation, which is applied in many recent publications \cite{mazzilli2021benefit}. GA is an evolutionary algorithm that tries to emulate Darwin’s theory of natural evolution. 
Due to the high number of generations required, GA can be very time-consuming when the cost function is costly to evaluate.

\begin{table}[t]
    \caption{Overview of the optimisation algorithm most used to tune the noise parameters of nonlinear Kalman filter.}
    \label{Table_Overview}
    \begin{center}
    \begin{tabular}{| >{\centering\arraybackslash}m{1in} | >{\centering\arraybackslash}m{1in} | >{\centering\arraybackslash}m{0.6in} |}
    \hline
    \textbf{Authors} & \textbf{Process noise tuning} & \textbf{Dataset}\\
    \hline
    van Aalst, 2018 \cite{van2018adaptive} & Manual Tuning & Experimental \\
    \hline
    Abbeel, 2005 \cite{abbeel2005discriminative} & Coordinate Ascent Algorithm & Experimental \\
    \hline
    Powell, 2002 \cite{powell2002automated} & Downhill Simplex Algorithm & Simulated \\
    \hline
    Mazzilli, 2021 \cite{mazzilli2021benefit}, Oshman, 2000 \cite{oshman2000optimal} & Multi-Objective GA, GA & Experimental \\
    \hline
    Acosta, 2019 \cite{acosta2019optimized} & FFO & Simulated \\
    \hline
    Heidfeld, 2021 \cite{heidfeld2021optimization} & Simulated Annealing & Experimental \\
    \hline
    Tang, 2021 \cite{tang2021reinforcement} & RL & Experimental \\
    \hline
    Escoriza, 2021 \cite{Escoriza2021} & Recurrent Neural Network & Experimental \\
    \hline
    Chen, 2018 \cite{chen2018weak}, Chen, 2019 \cite{chen2019kalman} & BO based on Gaussian \& t-Student Process & Simulated \\
    \hline
    \textbf{Current Paper} & \textbf{TSBO with t-Student Process} & \textbf{Experimental} \\
    \hline
    \end{tabular}
    \end{center}
\end{table}

\emph{Data-driven} optimisation algorithms such as BO are based on the creation of a surrogate model, Gaussian Processes GP \cite{chen2018weak}, or tSP \cite{chen2019kalman}, using as inputs the tuning parameters and as output the value of the cost function. The surrogate model does not only approximate the cost function, but also it is associated with a model probability distribution. The acquisition function takes this information to find the new parameters with the highest probability of being the best new optimum. Standard BO fits the surrogate model on parameters randomly taken from the sample space, but this can lead to a non-optimal surrogate model. Furthermore, it will require many cost function evaluations if the sample space is broad. 
Other data-driven approaches can be implemented; for instance, Reinforcement Learning RL can be applied to choose the parameters of the process noise, or RL can compensate for the error of the Kalman filter \cite{tang2021reinforcement} due to not-optimal process noise parameters. A similar concept is the so-called "KalmanNet" \cite{Escoriza2021} which uses a Recurrent Neural Network RNN to compute the Kalman gain, reducing the problems of non-optimal Kalman tuning. Despite the potential of RL and RNN, the amount of data required to train the Neural Networks makes them impractical for the tuning/improvement of Kalman filter for sideslip angle estimation. \looseness = -1

The main contributions of this paper are twofold. The first is the development of a new TSBO based on a tSP for tuning the process noise parameters of UKF for vehicle sideslip angle estimation. The TSBO in this paper differs from a recent TBSO related to electronic design \cite{torun2018global} because the surrogate model is based on a tSP to increase its robustness and ability to reach a better global optimum without increasing the required computational power.

The second contribution is that the developed TSBO reduces the number of simulations required to tune the Kalman filter with respect to the state-of-the-art GA \cite{mazzilli2021benefit,acosta2019optimized} and it improves the accuracy of the tuning respect GA. The tuning performance is tested and validated using an experimental dataset for vehicle  sideslip angle estimation.

\section{BAYESIAN OPTIMISATION}

Filter tuning consists of minimising a nonlinear, non-convex and non-smooth cost function due to the reasons already expressed in section I. A minimisation problem is:

\begin{equation}
    \begin{aligned}
        & \text{min}_{\;q\;\in\;Q}\;J\left(q\right) \\
        & \text{where} \;\; Q \subseteq\;\Re^d
    \end{aligned}
    \label{eq1}
\end{equation}

\noindent
where $q$ is the parameter vector of dimension $d$, $J\left(q\right)$ is the cost function, and $Q \subseteq\;\Re^d$ is the solution space. The function $J\left(q\right)$ is a "black-box" function because it is not accessible, but its outputs are observed based on some given inputs. Thus, "black-box" stochastic optimisation algorithms are considered. It is time-consuming to perform a dense sampling of the solution space $Q$, so the algorithm should focus on sampling the subset of $Q$ with the highest probability to contain the cost function optimum $q^*$. BO can deal with incomplete and sparse knowledge of the solution thanks to its probabilistic approach. It is based on the Bayes' theorem:

\begin{equation}
    P\left(\text{beliefs}\; |\; \text{data}\right) = \dfrac{P\left(\text{beliefs}\right) \times P\left(\text{data}\; |\; \text{beliefs}\right) }{P\left(\text{data}\right) }
    \label{eq2}
\end{equation}

\noindent
where data are the available observations of the function $J\left(q\right)$, beliefs are the beliefs of the shape of $J\left(q\right)$, and $P\left(\text{beliefs}\; |\; \text{data}\right)$, $P\left(\text{data}\; |\; \text{beliefs}\right)$, $P\left(\text{beliefs}\right)$, and $P\left(\text{data}\right) $ are the posterior, likelihood, and the marginal probabilities respectively. Thus, BO aims to find $q^*$ learning the shape of $J\left(q\right)$ through Bayesian inference. The TSBO overcomes the standard BO due to its ability to define a better ratio between exploration (function shape learning) and exploitation (approaching the optimum).

BO can be split into two steps: creating a stochastic process of the cost function, called "Surrogate Model", and building an "Acquisition Function" that uses the surrogate model to approach the optimum $q^*$.

\subsection{Surrogate Model}

\begin{figure}[t]
   \centering
   \setlength{\fboxrule}{0pt}
   \framebox{\parbox{3in}{\includegraphics[scale=0.28]{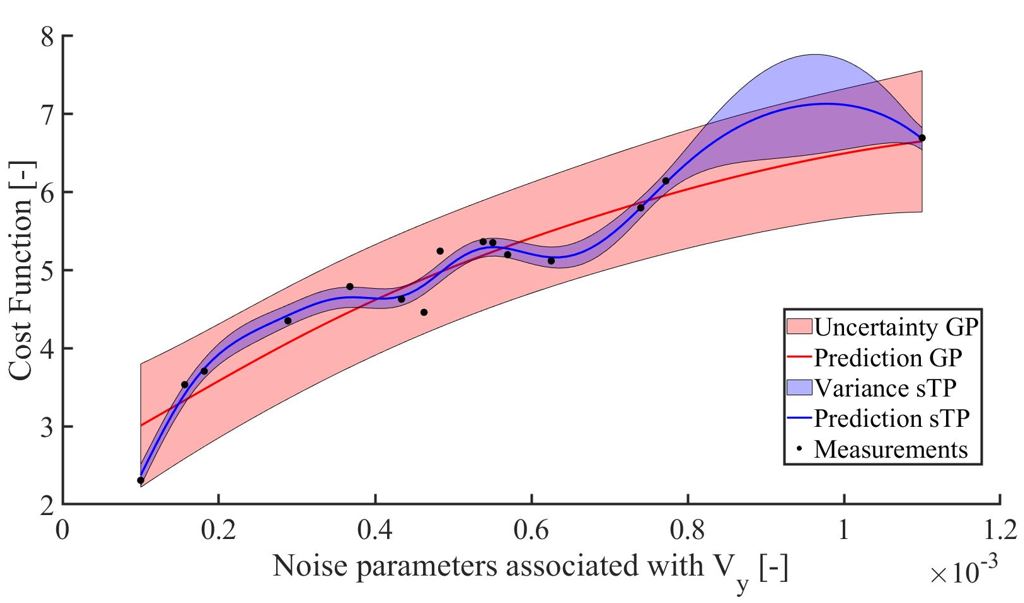}}}
    \caption{GP and sTP model of the cost function varying the process noise parameter related with the vehicle lateral velocity. The tSP allows more outliers and it can follow the measurements better than the GP. The shadow area is the variance of the tSP and the uncertainty for the GP.}
    \label{tSPvsGP}
\end{figure}

BO approximates the "black-box" cost function through a stochastic process called the surrogate model. It represents a probabilistic prior over the space of functions. The prior is updated as soon as the cost function is evaluated thanks to the Bayes' Theorem. Different priors are eligible for being the surrogate model; a partial list comprises: GP, tSP, Bayesian neural networks and polynomial chaos expansion. GP are the current state-of-art because it is a non-parametric model \cite{torun2018global}, and the Bayesian update step is analytical. Despite this, in this paper a tSP prior is selected for the surrogate model due to two reasons. The first one is that the t-Student distribution allows defining the level of Kurtosis, so it allows much more likely outliers \cite{shah2014student,tracey2018upgrading} than GP. The second reason is that the observations directly influence the predictive covariance \cite{shah2014student,tracey2018upgrading}. Both properties are beneficial for the tuning of UKF because the cost function can have discontinuities, for instance, due to the noise. Figure \ref{tSPvsGP} shows the explained differences between tSP and GP modelling in an example that models the cost function varying the process noise parameter related with lateral velocity. It is visible how tSP reduces the influence of the outliers, allowing the mean to have a better fitting. 

The tSP is defined as:

\begin{equation}
    f\left(q\right) = tSP\left(m\left(q\right),\;k\left(q,q'\right),\; \nu \right)
    \label{eq3}
\end{equation}

\noindent
where $m\left(q\right)$ represents the mean which is assumed equal to constant zero mean function because there is no prior \cite{torun2018bayesian}, $\nu$ are the degrees of freedom of the t-Student distribution, which defines the level of Kurtosis, and $k\left(q,q'\right)$ is the Kernel function between the inputs pair $q$ and $q'$. The choice of the Kernel is essential to have a reliable model fitting. This paper chooses the automatic relevance determination Matérn $5/2$ function, due to its generalisation capabilities and interpretability properties:

\begin{equation}
    k_{q,\;q'} = k(q,\;q')=\sigma_f^2\left(1+\frac{\sqrt{5}r}{\sigma_d}+\frac{5\,r^2}{3\,\sigma_d^2}\right)e^{-\frac{\sqrt{5}r}{\sigma_d}}
    \label{eq4}
\end{equation}

\noindent
where $r$ is the Euclidean distance between the two points $q$ and $q'$ calculated as:
\begin{equation}
    r = \sqrt{\left(q-q'\right)^T\left(q-q'\right)}
    \label{eq5}
\end{equation}

\noindent
where $\sigma_f$ and $\sigma_d$ are two hyperparamters, which are trained to minimize the negative log marginal likelihood of the tSP. It is essential to highlight that tSP should be accurate only when there are higher chances of finding the cost function optimum. Thus, every time a new observation is available, the surrogate model is retrained. tSP are implemented using the GPML Matlab Code version 4.2 \cite{williams2006gaussian}.

\subsection{Acquisition Function}

The Acquisition Function AF is responsible for moving the BO towards the optimum region, choosing the next sample point to be evaluated. Thus, it is responsible for defining the ratio between exploration, moving the search towards the area where the surrogate model is uncertain, and exploiting, moving the search towards the area with a higher chance of improving the current optimum. Ideally, the more focus is given to the exploration stage, the higher the chances to find the global optimum; however, it increases the computational time due to a larger number of cost function evaluations.

The next point to be evaluated ($q_{t+1}$) is usually obtained through an additional optimisation of the AF. This paper evaluates the AF on some candidate points obtained through the process that will be explained in Section IV to avoid further optimisation. The AF is evaluated on the candidate points \cite{williams2006gaussian, tracey2018upgrading} according to: 

\begin{equation}
    \hat{\mu}_{\,tSP}\left(q_{t+1}\right) = k_{q,\,q}^T\,K_{\tilde{q},\;\tilde{q}}^{-1}\,f\left(\tilde{q}\right)
    \label{eq6}
\end{equation}

\noindent
where $\tilde{q}$ are the sampling points already evaluated, $\hat{\mu}_{\,tSP}$ is the posterior mean of a tSP for a new sampling point $q$ and $k$ is defined in Eq. \ref{eq4}.

\begin{multline}
     \hat{\sigma}^2_{\,tSP}\left(q_{t+1}\right) = \left(\frac{\nu+f\left(\tilde{q}\right)^T\,K_{\tilde{q},\;\tilde{q}}^{-1}\,J\left(\tilde{q}\right)-2}{\nu + |D| -2}\right)\times(k\left(q,q\right) + \\-k_{q,\;\tilde{q}}^T\,K_{\tilde{q},\;\tilde{q}}^{-1}\,k_{q,\;\tilde{q}})
    \label{eq7}
\end{multline}

\noindent
where $\hat{\sigma}_{\,tSP}$ is the posterior covariance, and $D$ is the set of samples $D=\left[\left(\tilde{q}_1,J\left(\tilde{q}_1\right)\right),\; \left(\tilde{q}_2,J\left(\tilde{q}_2\right)\right), \; \left(\tilde{q}_3,J\left(\tilde{q}_3\right)\right),\;...\,\right]$. \newline

Two AFs are selected: the Expected Improvement (EI) and the Confidence Bound Minimisation (CBM) \cite{clare2020expected}.

The EI is defined as:

\begin{multline}
    EI_{\,tSP}\left(q\right)= \hat{\sigma}_{\,tSP}\left(q\right)\,\left(\frac{\nu}{\nu-1}\right)\,\left(1+\frac{z_s^2}{\nu}\right)\,\phi\left(z_s\right)+\\
    + \left[ \hat{y} - \hat{\mu}_{\,tSP}\right]\,\Phi\left(z_s\right)
    \label{eq8}
\end{multline}

\noindent
where $\phi\left(z_s\right)$ and $\Phi\left(z_s\right)$ are the probability density function and the cumulative distribution function of an univariate standard t-Student's random variable, $z_s$,  and $\hat{y}$ is the current optimum sampled by the BO. 

The CBM is formulated as:  

\begin{equation}
    CBM_{\,tSP}\left(q\right)= \hat{\sigma}_{\,tSP}\left(q\right)\,\sqrt{\beta} + \left[f^*- \hat{\mu}_{\,tSP}\right]
    \label{eq9}
\end{equation}

\noindent
where the $\beta$ defines the ratio between exploitation and exploration, and $f^*$ is the best known optimum, representing the prior knowledge of the optimum.

\section{TWO-STAGE BAYESIAN OPTIMISATION}

This paper proposes an evolution of the BO to tune the process noise parameters of the UKF. It is called TSBO, code by \cite{torun2018global}, and aims to reduce the number of function evaluations, and improve the ratio between exploitation and exploration. The trade-off improvement is obtained thanks to the subdivisions of the BO in two stages. The first, "fast exploration", aims to find the sample space region where the optimum is contained. The second, "pure exploitation", aims to fine-tune the restricted sample space obtained from the first stage. The first stage allows starting the surrogate model training from a single evaluation at the centre of the sample space. This allows the user not to evaluate a random set of process noises to fit the first surrogate model.

\subsection{Fast Exploration}

The fast exploration aims to narrow the sample space $Q\subseteq \Re^d$ where the BO search the optimum as fast as possible. It consists of subdividing the sample space into $2^d$ hyper-rectangles.  Every hyper-rectangle centre point is a candidate point, which the AF evaluates to choose the next sampling point $q_{\,t+1}$ where the cost function is evaluated. The following sampling point is not obtained as in the standard BO through an auxiliary optimisation of the AF, but it is the candidate point with the maximum value of the AF between the other candidate points. The cost function is only evaluated at points $q_{\,t+1}$. A new set of hyper-rectangles is generated starting from the hyper-rectangle enclosing the best candidate point. The two available AFs, EI and CBM, are alternatively used to evaluate the candidate points. After a predefined number of iterations $MAX_{AF}$, the AF which would have had the greatest gain to the optimisation is selected as final AF.

The fast exploration stage is repeated until the euclidean distance between the new best sampling point $q_{t+1,\;max}$ and the previous best sampling point $q_{t\;max}$ is below a user-defined threshold $TR_{FE}$ for a number of times $MAX_{FE}$:

\begin{equation}
    n = \begin{cases}
            n+1\,, \;\;\;\;\;\;\;& \text{if} \;\;\; ||q_{\,t+1,\,max} - q_{\,t,\,max}|| < TR_{FE}\\
            0\,, \;\;\;\;\;\;\; & \text{otherwise} \\
        \end{cases}
    \label{eq10}
\end{equation}

Every time the cost function is evaluated and a new measurement is available; the surrogate model is retrained to improve the fitting with the black-box function. 

Algorithm \ref{alg1} sums up the fast exploration stage of the TSBO.

\begin{algorithm}
    \caption{Fast Exploration stage of the TSBO}
    \label{alg1}
    \begin{algorithmic}
        \Require A sample space $Q$ and an evaluated sampling point at the center of $Q$
        \State Fit the tSP Surrogate model
        \State Subdivide $Q$ in $2^d$ hyper-rectangles
        \While{$n \le MAX_{FE}$}
            \State Compute candidate points at the center of hyper-rectangles
            \State Evaluate AF at the candidate points
            \State Pick the candidate points with the maximum AF
            \If{$||q_{\,t+1\,max} - q_{\,t\,max}|| < TR_{FE}$}
                \State $n+1 \gets n$
            \ElsIf{$||q_{\,t+1\,max} - q_{\,t\,max}|| \geq TR_{FE}$}
                \State $n \gets n$
            \EndIf
        \State Evaluate the cost function in the sampling points
        \State Train the tSP of the Surrogate with the new observation
        \State Subdivide the hyper-rectangle which were enclosing the best candidate points in $2^d$ hyper-rectangles 
        \EndWhile
    \end{algorithmic}
\end{algorithm}

\begin{algorithm}
    \caption{Pure Exploitation stage of the TSBO}
    \label{alg2}
    \begin{algorithmic}
        \Require $n = MAX_{FE}$
        \While{$n_{iter} \le MAX_{PE}$, and the output from Algorithm 1}
            \State Compute candidate points at the center of hyper-rectangles
            \State Evaluate AF at the candidate points
            \State Pick the candidate points with the maximum AF
            \State Evaluate the cost function in the sampling points
            \If{$n_{iter} \le MAX_{SM}$}
                \State Train the tSP of the Surrogate with the new observation
            \EndIf
        \State Subdivide the hyper-rectangle which were enclosing the best candidate points in $3^d$ hyper-rectangles 
        \EndWhile
    \end{algorithmic}
\end{algorithm}

\subsection{Pure Exploitation}

The pure exploitation stage follows a procedure similar to the fast exploration, but it specialises in refining the optimum position. The first step is the computation of the sample space $\tilde{Q}^d$ around the optimum computed in the fast exploration stage:

\begin{equation}
    \tilde{Q}^d = \begin{bmatrix}
                        \left(1-\alpha\right)\,q^*_{\,1} & \left(1+\alpha\right)\,q^*_{\,1}\\
                        \vdots & \vdots\\
                        \left(1-\alpha\right)\,q^*_{\,d} & \left(1+\alpha\right)\,q^*_{\,d}
                    \end{bmatrix}
    \label{eq11}
\end{equation}

\noindent
where $\alpha$ is a hyper-parameter that defines how wide the sample space should be, and $q^*_i$ is the optimum sampling point obtained in the fast exploration stage. The new sample space $\tilde{Q}^d$ is divided into $3^d$ hyper-rectangles to have more candidate points to evaluate. The higher number of candidate points helps the pure exploitation stage to converge faster to the global optimum of the function. The rest of the procedure is the same of the fast exploration. The only difference is that when the number of evaluated sampling points $n_{iter}$ is higher than a predefined threshold $MAX_{SM}$, the tSP is no more retrained.

Algorithm \ref{alg2} summarises the pure exploitation stage of the TSBO, which is iterated until $n_{iter}$ is lower than a user-defined threshold $MAX_{PE}$.

\section{APPLICATION: UKF TUNING}

The proposed TSBO based on tSP is tested on the tuning of the process noise parameters of a UKF for vehicle sideslip angle estimation. In this paper, the single-track model with tyre axle forces computed by the Dugoff tyre is chosen as the vehicle model. The states are the longitudinal velocity at the Center-of-Gravity CoG $V_x$, the lateral velocity at the CoG $V_y$ and the yaw rate $\dot{\psi}$ while the vehicle measurements are the $V_x$, the lateral acceleration at the CoG $a_y$ and the $\dot{\psi}$. The vehicle sideslip angle $\beta_s$ is obtained as $\beta_s = \arctan{\frac{V_y}{V_x}}$:

\subsection{Dataset}

All the manoeuvres that compose the training set and test set are taken from a real-world experimental dataset composed of 216 manoeuvres which correspond to 2 hrs of driving. It puts together standard vehicle dynamics manoeuvres, e.g. double lane change, slalom, random steer, J-turn, spiral, braking in a turn, and steady-state circular tests, together with recorded laps at the Papenburg track. All manoeuvres were performed on dry asphalt with tyres inflated according to the vehicle specifications.

All manoeuvres are recorded using the BMW test platform instrumented by the inertial measurement unit, wheel force transducers for all four wheels, GPS and an optical sensor from Corrsys-Datron to measure the sideslip angle. The high-end optical speed sensor is not present in consumer vehicles and will be used as a reference for TBSO tuning and validation.

The UKF should be tuned for different working conditions, so various vehicle maneuvers are jointly considered in the training set. The training set is composed of 8 manoeuvres: 1 braking in a turn, 1 skidpad, 2 J-turn, 2 slaloms and 2 lance change.

A test set composed of 23 manoeuvres is taken to prove the generalization capability of the approach. All 23 maneuvers of the test set are different from the training set even if, some of them have similar characteristics. The manoeuvres are: 2 braking in a turn, 2 skidpad, 5 J-turn, 4 slalom, 4 lane change, 2 random steer, 1 lap track and 3 spiral. The characteristics of the manoeuvres cover a wide range of vehicle driving behaviours and different settings of the Electronic Stability Control.

\subsection{Cost Function Formulation}

In this paper, the UKF tuning aims to improve the accuracy of the vehicle state estimation, therefore, this paper focuses on an objective function defined by minimising the Normalised Root Mean Squared Error NRMSE rather than maximising the negative log-likelihood \cite{abbeel2005discriminative}. 

The minimisation problem is described as:

\begin{equation}
    \begin{aligned}
        & \text{min}_{\;q\;\in\;Q} \;J\left(q\right) \\
        & \text{where} \;\; Q =\;\begin{bmatrix}
                                        q_{\,1,\, min} & q_{\,1,\, max}\\
                                        \vdots & \vdots\\
                                        q_{\,d,\, min} & q_{\,d,\, max}
                                    \end{bmatrix}
    \end{aligned}
    \label{eq13}
\end{equation}

\noindent
where $J$ is the cost function defined as:

\begin{equation}
    \begin{aligned}
        & J = W_1\,E_{\beta} + W_2\,E_{\dot{\psi}} + W_3\,E_{a_y} \\
        &\text{where}\;\; E_j = \sqrt{\frac{\sum_{i=1}^{N{\,man}}\left(NRMSE_{j,\;i}\right)^2\,N_{s,\;i}}{\sum_{i=1}^{N_{\,man}}N_{s,\;i}}} 
    \end{aligned}
    \label{eq14}
\end{equation}

\noindent
where $W_1$, $W_2$, $W_3$ are the weights of the cost function, $N_s$ is the length of the manoeuvre measured in sampling points, $N_{man}$ is the number of manoeuvres considered in the training set, and $NRMSE$ is computed as follow:.

\begin{equation}
    NRMSE = \sqrt{\frac{\sum_{k=1}^{N_s}\left(\hat{X}_k-X_k\right)^2}{N_s}}\frac{1}{\text{max}\left(|X|\right)}
    \label{eq15}
\end{equation}

\noindent
where $X$ and $\hat{X}$ are the vectors of the measured and estimated states respectively.

\subsection{Key Performance Indicator}

The accuracy of the estimation of sideslip angle is evaluated through the use of the root mean squared error RMSE, maximum absolute error MAE, the RMSE and the MAE when the absolute vehicle acceleration $a_y$ is greater than \SI{4}{m/s^2}, which are respectively abbreviated with RMSE\textsubscript{NON} and MAE\textsubscript{NON}. The latter is used to analyse how the UKF tuning is especially relevant when the vehicle behaves in a nonlinear way.

\section{RESULTS}

 The performance in terms of minimisation time and optimum localisation of the TSBO based on tSP is compared with the current state-of-the-art technique GA and with a TSBO based on GP. All optimisations perform in the same sample space $Q$ where all the process noise parameters for $V_x$, $V_y$ and $\dot{\psi}$ vary from $10^{-10}$ to 1.

The optimisation parameters for TSBO based on tSP and GP are summarised in Table \ref{par_TSBO}. The parameters for GA are summarized in Table \ref{par_GA}. 

\begin{table}[b]
    \caption{User-defined parameters for TSBO optimisations.}
    \label{par_TSBO}
    \begin{center}
    \begin{tabular}{| >{\centering\arraybackslash}m{0.9in} | >{\centering\arraybackslash}m{0.7in} | >{\centering\arraybackslash}m{0.7in} |}
    \hline
    \textbf{Parameters} & \textbf{TSBO - tSP} & \textbf{TSBO - GP} \\
    \hline
    Likelihood function  & t-Student ’s & Gaussian \\
    \hline
    Inference  & Variational Bayes & Exact \\
    \hline
    Kernel  function & Matérn ARD $5/2$ & Matérn ARD $5/2$ \\
    \hline
    Mean function  & Constant & Constant \\
    \hline
    $\nu$  & 15 & NA \\
    \hline
    $MAX_{FE}$  & 15 & 15 \\
    \hline
    $MAX_{PE}$  & 40 & 40 \\
    \hline
    $MAX_{SM}$  & 38 & 38 \\
    \hline
    $TR_{FE}$  & $0.01\parallel \;q^* \parallel$ & $0.01\parallel \;q^* \parallel$ \\
    \hline
    $\beta$  & 0.01 & 0.01 \\
    \hline
    $\alpha$  & 0.15 & 0.15 \\
    \hline
    $W_1$, $W_2$, $W_3$  & 5, 1, 1 & 5, 1, 1\\
    \hline
    \end{tabular}
    \end{center}
\end{table}

\begin{table}[t]
    \caption{User-defined parameters for GA optimisation.}
    \label{par_GA}
    \begin{center}
    \begin{tabular}{| >{\centering\arraybackslash}m{0.9in} | >{\centering\arraybackslash}m{0.9in} |}
    \hline
    \textbf{Parameters} & \textbf{GA} \\
    \hline
    Population size  & 15 \\
    \hline
    Max generations  & 15\\
    \hline
    Élite count & 0.75 \\
    \hline
    Crossover fraction & 0.8 \\
    \hline
    $W_1$, $W_2$, $W_3$  & 5, 1, 1 \\
    \hline
    \end{tabular}
    \end{center}
\end{table}

\begin{figure}[b]
   \centering
   \setlength{\fboxrule}{0pt}
   \framebox{\parbox{3in}{\includegraphics[scale=0.3]{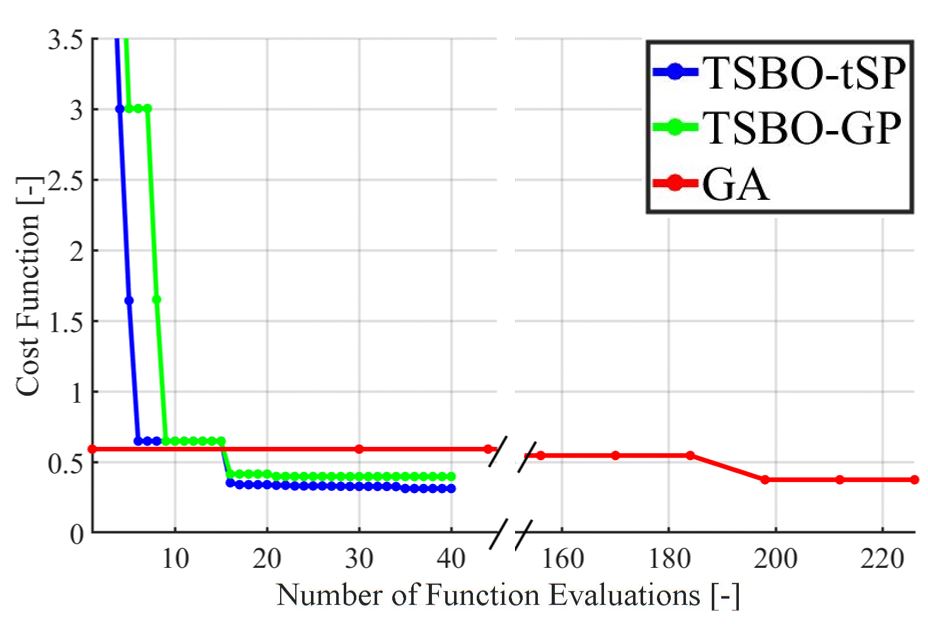}}}
    \caption{Performance comparison of TSBO based on tSP, TSBO based on GP and GA.}
    \label{res_com_pic}
\end{figure}

The optimum cost and optimisation time for the TSBO based on tSP, TSBO based on GP and GA are summarised in Table \ref{res_com_tab} and Figure \ref{res_com_pic}.

\begin{table}[t]
    \caption{Numerical results of the optimum cost function and optimisation time using different algorithms for UKF tuning.}
    \label{res_com_tab}
    \begin{center}
    \begin{tabular}{| >{\centering\arraybackslash}m{0.9in} | >{\centering\arraybackslash}m{0.9in} | >{\centering\arraybackslash}m{0.9in} |}
    \hline
    \textbf{Algorithm} &  \textbf{Total cost [-] }  &  \textbf{Optim. time [s]}\\
    \hline        
    Manual tuning &   1.714             & NA\\ 
    \hline
    GA            &   0.378             & 1599\\
    \hline
    TSBO - GP     &   0.400             & \textbf{306}\\
    \hline
    TSBO - tSP    &   \textbf{0.316}    & 320\\ 
    \hline
    \end{tabular}
    \end{center}
\end{table}

Figure \ref{res_com_pic} shows how both TSBOs approach the global optimum faster than GA. Where TSBO converges after 40 evaluations of the cost function, the GA needs at least 220 evaluations, resulting in TSBO being on average \SI{79}{\%} faster than GA. Thus, TSBOs are particularly convenient when the cost function evaluation is computationally heavy, for instance, due to numerous manoeuvres considered in the training set. Despite the faster optimisation, TSBO based on tSP reaches an optimum \SI{16.4}{\%} lower than the GA optimum. The proposed algorithm outperforms the state-of-the-art approach both in terms of optimisation time and optimisation accuracy. TSBO based on GP is \SI{4.4}{\%} faster than TSBO based on tSP. The rise in optimisation time is due to the surrogate model's training phase, which is slower for the tSP due to the more optimisation parameters. The better optimisation accuracy, \SI{21}{\%}, of the tSP than the GP is due to the higher Kurtosis of the t-Student's distribution which makes the non-parametric model capable of dealing with outliers. A much higher optimisation accuracy compensates for the slight higher computational time. \newline

Table \ref{Trai} presents the RMSE, MAE, RMSE\textsubscript{NON} and MAE\textsubscript{NON} of the sideslip angle for the various optimised UKFs. The reported values are the average of all manoeuvres composing the training set. Table \ref{Trai} highlights how the cost function's minimisation is related to the KPIs used to compare the estimation accuracy because TSBO based on tSP outperforms the other algorithms. On average, the RMSE of the sideslip angle optimised with tSBO based on tSP is \SI{15.4}{\%} lower than the one with GA, and the MAE is \SI{15}{\%} lower. Particularly important is the improvement of the RMSE\textsubscript{NON} and of the MAE\textsubscript{NON} because when the $|a_y| \ge$ \SI{4}{m/s^2} the sideslip angle estimation becomes more critical for vehicle stability control. The UKF optimised by the TSBO based on tSP reduces the RMSE\textsubscript{NON}, and the MAE\textsubscript{NON} of the UKF optimised by GA of \SI{24.6}{\%} and \SI{12.5}{\%}, respectively. The accuracy of the UKF tuned by GA is higher than the one optimised by TSBO based on GP.

Table \ref{Vali} reports the same information as Table \ref{Trai}, but for the test set, to check that the numerical optimisation algorithms do not overfit the training set leading to worse performance in other manoeuvres. Table \ref{Vali} shows that TSBO based on tSP still behaves better than all other algorithms for the 4 KPIs, proving better performance in tuning the process noise UKF parameters.
On average, the RMSE of the sideslip angle optimised with tSBO based on tSP is \SI{9.9}{\%} lower than the one with GA, and the MAE is \SI{17.6}{\%} lower. The improvement of the RMSE\textsubscript{NON} and of the MAE\textsubscript{NON} is particularly significant because when the $|a_y| \ge$ \SI{4}{m/s^2}, the sideslip angle estimation becomes more critical for vehicle stability control. The UKF optimised with the TSBO based on tSP reduces the RMSE\textsubscript{NON}, and the MAE\textsubscript{NON} of the UKF optimised with GA of \SI{10.6}{\%} and \SI{9.8}{\%}, respectively. The improvement in the test set is lower than the one obtained in the training set but still particularly relevant, especially for what concerns the MAE. \newline

\begin{table}[t]
    \caption{Comparison of the average KPIs for the UKF tuned by manual tuning, GA, TSBO based on GP and TSBO based on sTP. Results of the training set.}
    \label{Trai}
    \begin{center}
    \begin{tabular}{| >{\centering\arraybackslash}m{0.5in} | >{\centering\arraybackslash}m{0.5in} | >{\centering\arraybackslash}m{0.5in} | >{\centering\arraybackslash}m{0.5in} | >{\centering\arraybackslash}m{0.5in} |}
    \hline
    \textbf{KPI} &  \textbf{Manual}  &  \textbf{GA} &  \textbf{TSBO - GP}&  \textbf{TSBO - tSP}\\
    \hline
    RMSE [deg]                        & 1.028  & 0.500 & 0.634  & \textbf{0.423} \\
    \hline
    MAE [deg]                         & 1.196  & 0.900 & 1.252  & \textbf{0.765} \\
    \hline
    RMSE\textsubscript{NON} [deg]     & 0.716  & 0.568 & 0.855  & \textbf{0.428} \\
    \hline
    MAE\textsubscript{NON} [deg]      & 0.818  & 0.671 & 0.961  & \textbf{0.587} \\
    \hline
    \end{tabular}
    \end{center}
\end{table}

\begin{table}[b]
    \caption{Comparison of the average KPIs for the UKF tuned by manual tuning, GA, TSBO based on GP and TSBO based on sTP. Results of the test set.}
    \label{Vali}
    \begin{center}
    \begin{tabular}{| >{\centering\arraybackslash}m{0.5in} | >{\centering\arraybackslash}m{0.5in} | >{\centering\arraybackslash}m{0.5in} | >{\centering\arraybackslash}m{0.5in} | >{\centering\arraybackslash}m{0.5in} |}
    \hline
    \textbf{KPI} &  \textbf{Manual}  &  \textbf{GA} &  \textbf{TSBO - GP}&  \textbf{TSBO - tSP}\\
    \hline
    RMSE [deg]                        & 0.577 & 0.383 & 0.431 & \textbf{0.345}  \\
    \hline
    RMSE [deg]                        & 0.672 & 0.529 & 0.689 & \textbf{0.436}  \\
    \hline
    RMSE\textsubscript{NON} [deg]     & 1.328  & 0.860 & 1.030  & \textbf{0.769}  \\
    \hline
    MAE\textsubscript{NON} [deg]      & 1.047  & 0.746 & 0.902 & \textbf{0.673} \\
    \hline
    \end{tabular}
    \end{center}
\end{table}

\begin{figure}[t]
   \centering
   \setlength{\fboxrule}{0pt}
   \framebox{\parbox{3in}{\includegraphics[scale=0.27]{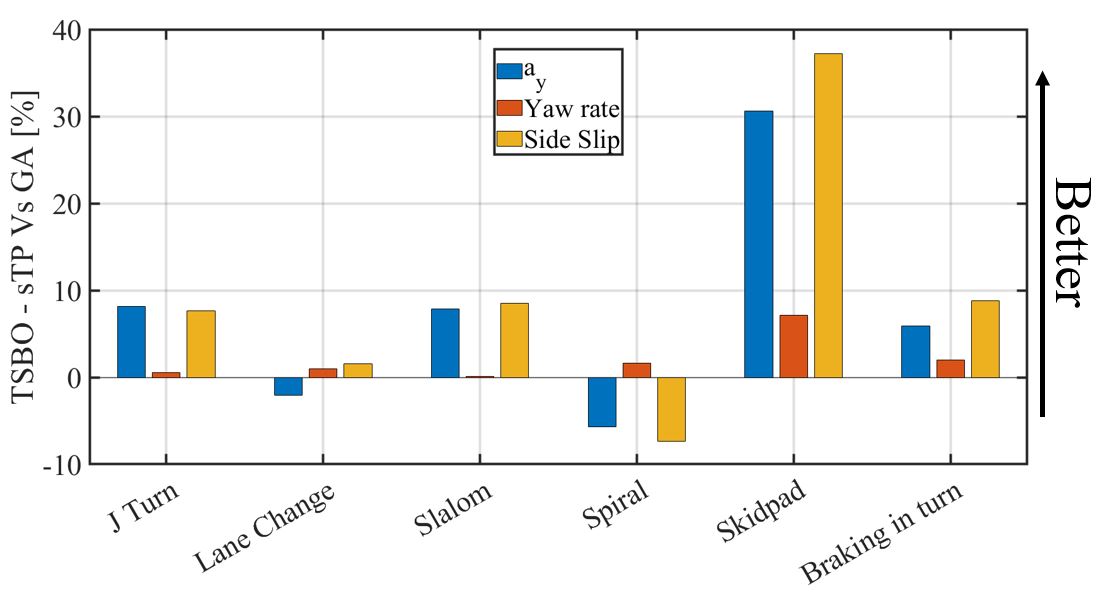}}}
    \caption{Relative improvement of the UKF tuned by TSBO based on tSP with respect to the UKF tuned by the state-of-the-art GA. The improvement is measured through the sideslip angle's RMSE for 6 different maneuver chosen from the test set.}
    \label{bar}
\end{figure}

\begin{figure}[t]
   \centering
   \setlength{\fboxrule}{0pt}
   \framebox{\parbox{3in}{\includegraphics[scale=0.12]{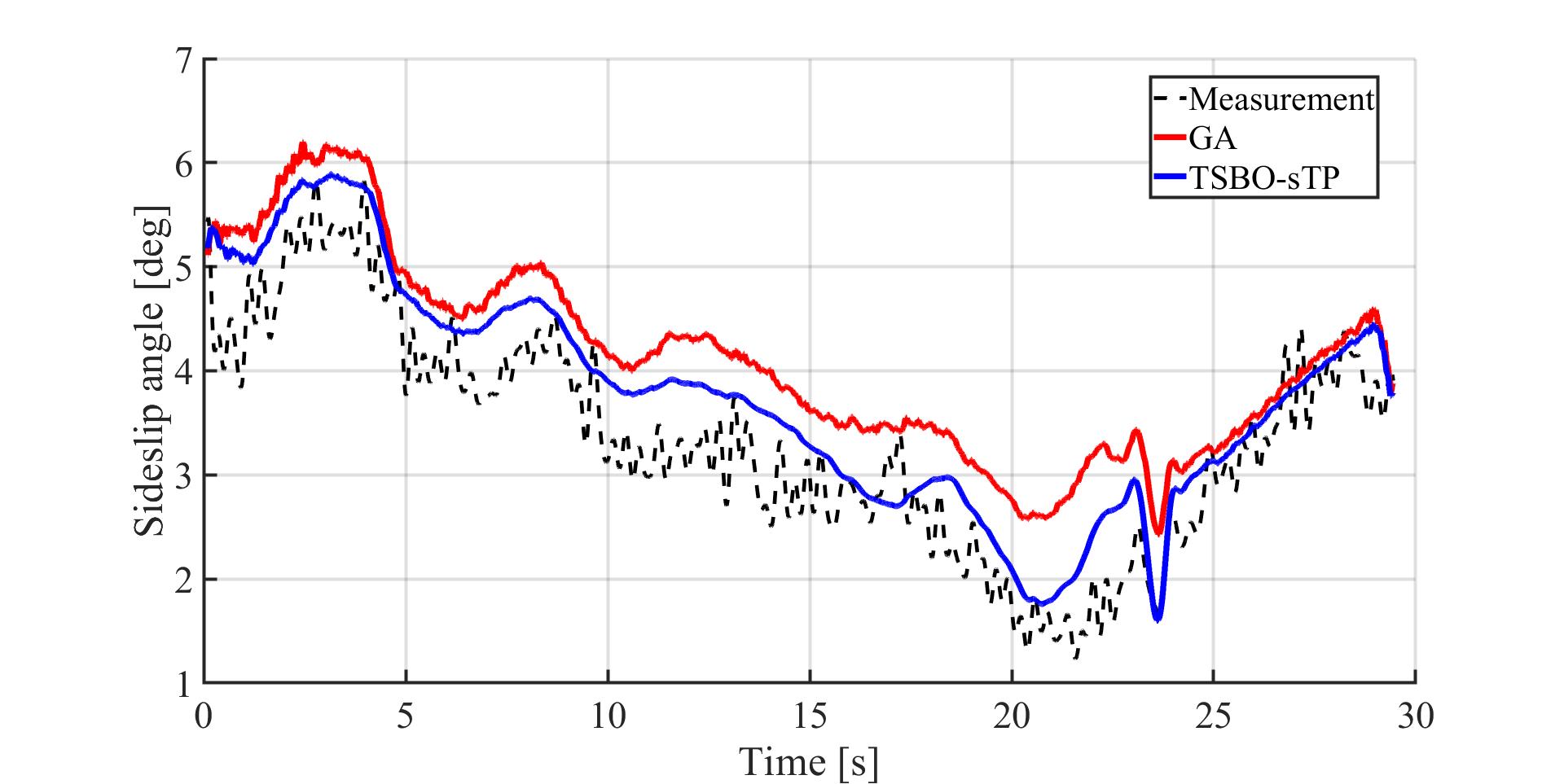}}}
    \caption{Sideslip angle estimation in a skidpad maneuver.}
    \label{SKIDPAD}
\end{figure}

\begin{figure}[t]
   \centering
   \setlength{\fboxrule}{0pt}
   \framebox{\parbox{3in}{\includegraphics[scale=0.12]{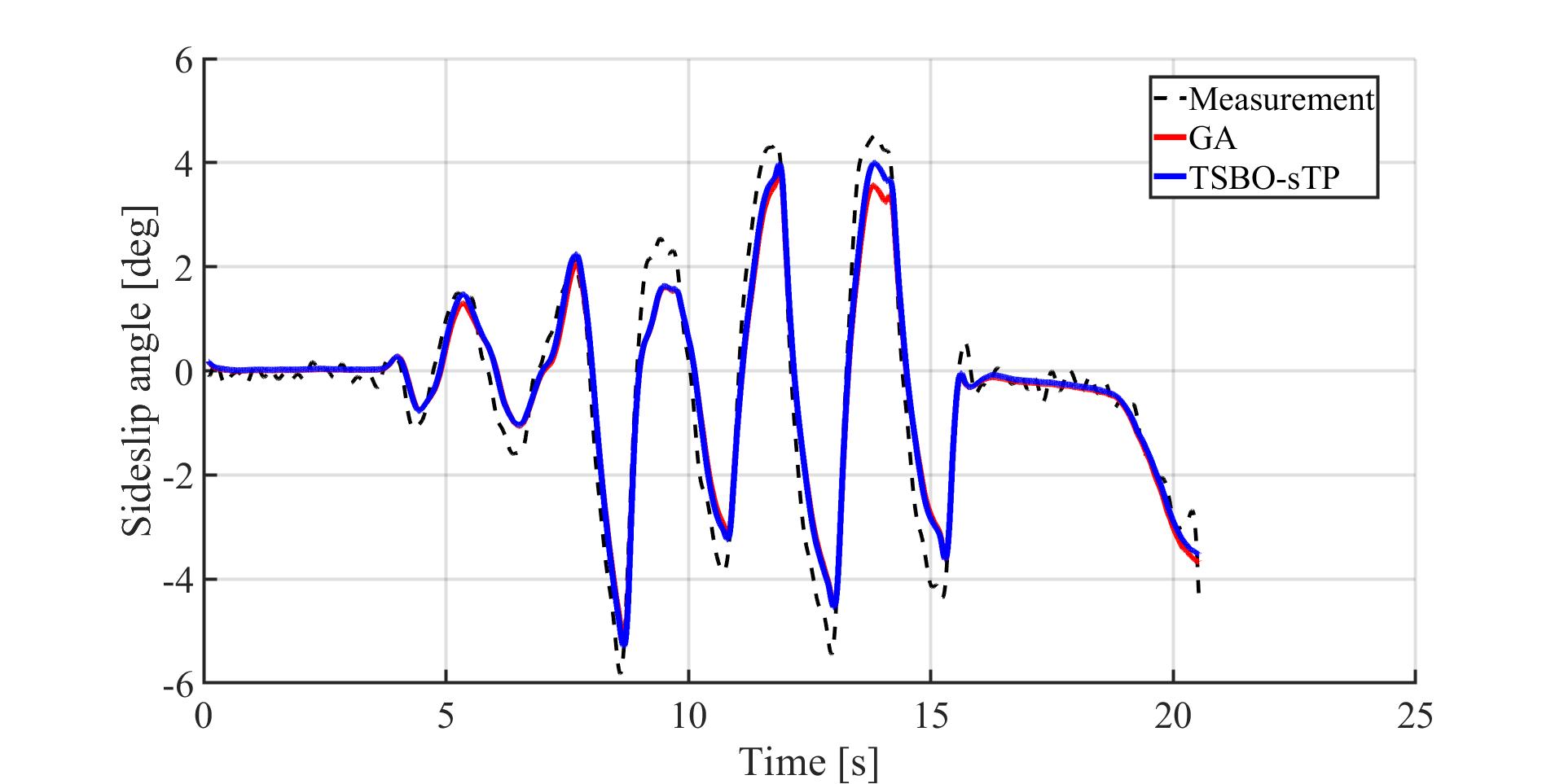}}}
    \caption{Sideslip angle estimation in a slalom maneuver.}
    \label{SLALOM}
\end{figure}

\begin{figure}[ht]
   \centering
   \setlength{\fboxrule}{0pt}
   \framebox{\parbox{3in}{\includegraphics[scale=0.12]{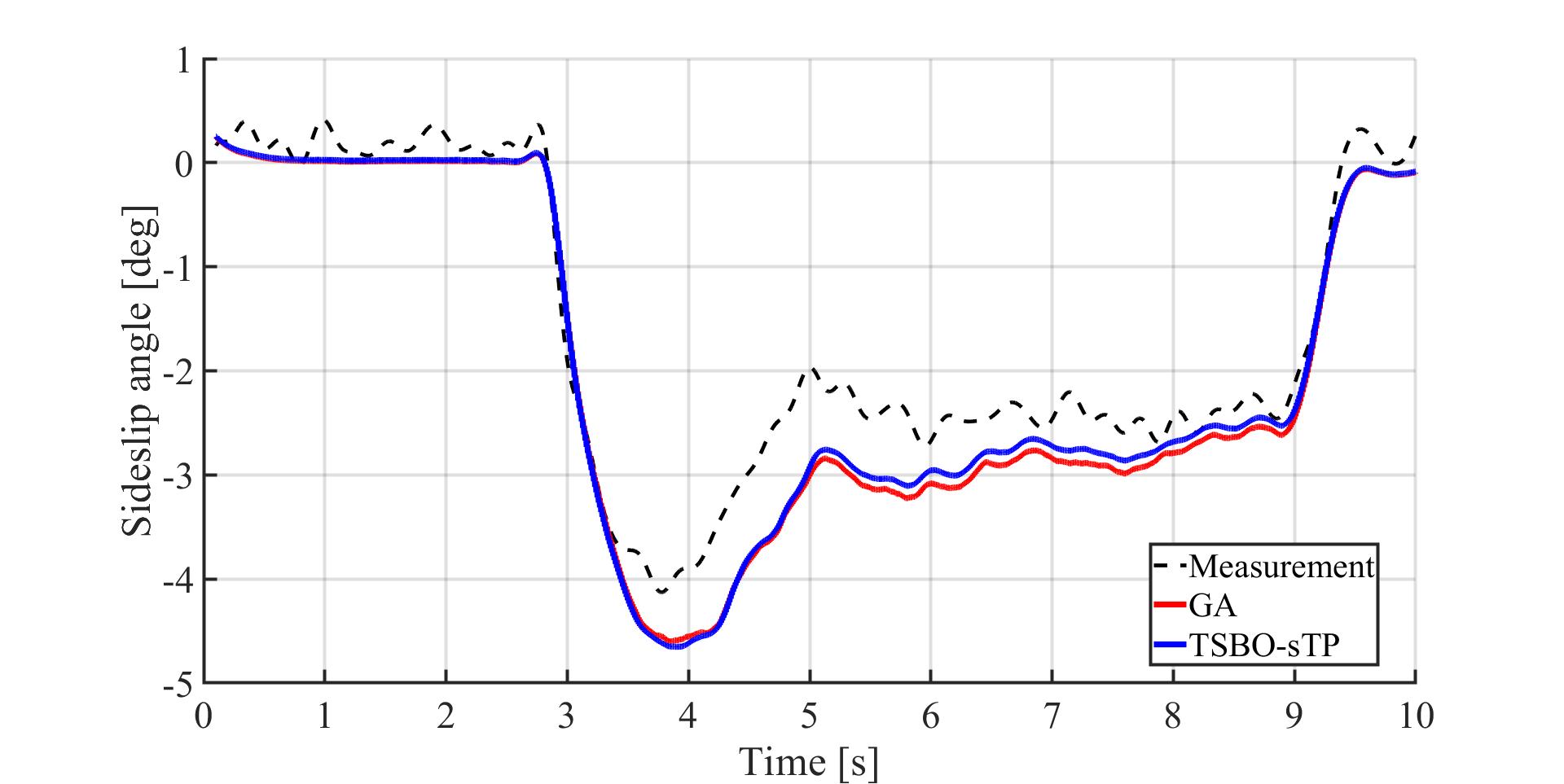}}}
    \caption{Sideslip angle estimation in a J-turn maneuver.}
    \label{JTURN}
\end{figure}

Figure \ref{bar} shows the relative improvement of the UKF tuned by TSBO based on tSP and GA in 6 different maneuvers of the test set. The relative improvement is measured with the RMSE of the sideslip angle, $a_y$ and $\dot{\psi}$. In 5 out of 6 maneuvers the TSBO reaches higher performance than the GA. The spiral maneuver is the only one which shows a degraded accuracy in the RMSE of sideslip angle  of \SI{8}{\%}. Performance degradation in this particular maneuver is considered acceptable due to the various driving conditions of the test set. The relative improvement is significant especially for the sideslip angle and for the $a_y$, while the $\dot{\psi}$ does not have this. The reason is that the $\dot{\psi}$ has lower maximum absolute error compared with the other variables, so the relative improvement is less noticeable. Figure \ref{bar} shows how TSBO can enhance the performance of the UKF for both transient (J-turn, slalom) and steady state (skidpad) maneuvers. Significant is the improvement in the skidpad, see also Figure \ref{SKIDPAD}. The reason is that during a steady-state maneuver the vehicle yaw acceleration is almost null so the difference between estimated and measured tyre forces becomes essential for the $V_y$ estimation. The best way to reduce the tyre force error is to optimise the process noise parameters.

Despite this, the improvement in sideslip angle estimation is noticeable also for a J-turn, see Figure \ref{JTURN}, and a slalom manoeuvre, see Figure \ref{SLALOM}.
Figure \ref{SLALOM} shows how the improvement in the tuning of the UKF is noticeable at the sideslip angle peaks of the slalom maneuver. The improvement at the peaks is more remarkable when the sideslip angle is above \SI{3}{deg} while when it is between 0 and \SI{3}{deg} the difference is not visible due to the lower absolute error of both UKFs.
Figure \ref{JTURN} shows the sideslip angle estimation in J-turn and it highlights another time how the UKF tuned by TSBO based on tSP reduces the estimation error especially when the sideslip angle is above \SI{4}{deg}.

\section{CONCLUSIONS}

This paper presented a two-stage Bayesian optimisation approach based on t-Student process to tune the process noise parameters of a unscented Kalman filter for sideslip angle estimation. The proposed approach reaches the cost function optimum \SI{79.9}{\%} faster than the state-of-the-art GA and the optimum is \SI{16.4}{\%} better than the one obtained by GA. The obtained results are tested on an experimental test set composed by transient and quasi-steady state maneuvers. Furthermore, two-stage Bayesian optimisation creates a physical representation of the black-box cost function thanks to surrogate model training. The performance of the two-stage Bayesian optimisation based on t-Student process is also compared with the two-stage Bayesian optimisation based on Gaussian process. The algorithm based on t-Student process is \SI{4.4}{\%} slower than the one based on Gaussian process but it reaches a \SI{21}{\%} better optimum. As future research, the plan is to tune not only the process noise parameters of the unscented Kalman filter but also the parameters defining the location of the sigma points.

\addtolength{\textheight}{-15cm}   


\bibliographystyle{IEEEtran}
\bibliography{IEEEabrv,references}

\begin{thebibliography}{10}
\providecommand{\url}[1]{#1}
\csname url@samestyle\endcsname
\providecommand{\newblock}{\relax}
\providecommand{\bibinfo}[2]{#2}
\providecommand{\BIBentrySTDinterwordspacing}{\spaceskip=0pt\relax}
\providecommand{\BIBentryALTinterwordstretchfactor}{4}
\providecommand{\BIBentryALTinterwordspacing}{\spaceskip=\fontdimen2\font plus
\BIBentryALTinterwordstretchfactor\fontdimen3\font minus
  \fontdimen4\font\relax}
\providecommand{\BIBforeignlanguage}[2]{{%
\expandafter\ifx\csname l@#1\endcsname\relax
\typeout{** WARNING: IEEEtran.bst: No hyphenation pattern has been}%
\typeout{** loaded for the language `#1'. Using the pattern for}%
\typeout{** the default language instead.}%
\else
\language=\csname l@#1\endcsname
\fi
#2}}
\providecommand{\BIBdecl}{\relax}
\BIBdecl

\bibitem{chen2018weak}
Z.~Chen, C.~Heckman, S.~Julier, and N.~Ahmed, ``Weak in the nees?: Auto-tuning
  kalman filters with bayesian optimization,'' in \emph{2018 21st International
  Conference on Information Fusion (FUSION)}.\hskip 1em plus 0.5em minus
  0.4em\relax IEEE, 2018, pp. 1072--1079.

\bibitem{abbeel2005discriminative}
P.~Abbeel, A.~Coates, M.~Montemerlo, A.~Y. Ng, and S.~Thrun, ``Discriminative
  training of kalman filters.'' in \emph{Robotics: Science and systems},
  vol.~2, 2005, p.~1.

\bibitem{chen2021time}
Z.~Chen, C.~Heckman, S.~Julier, and N.~Ahmed, ``Time dependence in kalman
  filter tuning,'' in \emph{2021 IEEE 24th International Conference on
  Information Fusion (FUSION)}.\hskip 1em plus 0.5em minus 0.4em\relax IEEE,
  2021, pp. 1--8.

\bibitem{acosta2019optimized}
M.~Acosta and S.~Kanarachos, ``Optimized vehicle dynamics virtual sensing using
  metaheuristic optimization and unscented kalman filter,'' in
  \emph{Evolutionary and Deterministic Methods for Design Optimization and
  Control With Applications to Industrial and Societal Problems}.\hskip 1em
  plus 0.5em minus 0.4em\relax Springer, 2019, pp. 275--290.

\bibitem{mazzilli2021benefit}
V.~Mazzilli, D.~Ivone, S.~De~Pinto, L.~Pascali, M.~Contrino, G.~Tarquinio,
  P.~Gruber, and A.~Sorniotti, ``On the benefit of smart tyre technology on
  vehicle state estimation,'' \emph{Vehicle System Dynamics}, pp. 1--26, 2021.

\bibitem{tang2021reinforcement}
Y.~Tang, L.~Hu, Q.~Zhang, and W.~Pan, ``Reinforcement learning compensated
  extended kalman filter for attitude estimation,'' in \emph{2021 IEEE/RSJ
  International Conference on Intelligent Robots and Systems (IROS)}.\hskip 1em
  plus 0.5em minus 0.4em\relax IEEE, 2021, pp. 6854--6859.

\bibitem{torun2018global}
H.~M. Torun, M.~Swaminathan, A.~K. Davis, and M.~L.~F. Bellaredj, ``A global
  bayesian optimization algorithm and its application to integrated system
  design,'' \emph{IEEE Transactions on Very Large Scale Integration (VLSI)
  Systems}, vol.~26, no.~4, pp. 792--802, 2018.

\bibitem{heidfeld2020ukf}
H.~Heidfeld, M.~Sch{\"u}nemann, and R.~Kasper, ``Ukf-based state and tire slip
  estimation for a 4wd electric vehicle,'' \emph{Vehicle System Dynamics},
  vol.~58, no.~10, pp. 1479--1496, 2020.

\bibitem{heidfeld2021optimization}
H.~Heidfeld and M.~Sch{\"u}nemann, ``Optimization-based tuning of a hybrid ukf
  state estimator with tire model adaption for an all wheel drive electric
  vehicle,'' \emph{Energies}, vol.~14, no.~5, p. 1396, 2021.

\bibitem{van2018adaptive}
S.~van Aalst, F.~Naets, B.~Boulkroune, W.~De~Nijs, and W.~Desmet, ``An adaptive
  vehicle sideslip estimator for reliable estimation in low and high excitation
  driving,'' \emph{IFAC-PapersOnLine}, vol.~51, no.~9, pp. 243--248, 2018.

\bibitem{kerst2019}
S.~Kerst, B.~Shyrokau, and E.~Holweg, ``A model-based approach for the
  estimation of bearing forces and moments using outer ring deformation,''
  \emph{IEEE Transactions on Industrial Electronics}, vol.~67, no.~1, pp.
  461--470, 2019.

\bibitem{powell2002automated}
T.~D. Powell, ``Automated tuning of an extended kalman filter using the
  downhill simplex algorithm,'' \emph{Journal of Guidance, Control, and
  Dynamics}, vol.~25, no.~5, pp. 901--908, 2002.

\bibitem{oshman2000optimal}
Y.~Oshman and I.~Shaviv, ``Optimal tuning of a kalman filter using genetic
  algorithms,'' in \emph{AIAA Guidance, Navigation, and Control Conference and
  Exhibit}, 2000, p. 4558.

\bibitem{Escoriza2021}
A.~L. Escoriza, G.~Revach, N.~Shlezinger, and R.~J.~G. van Sloun, ``Data-driven
  kalman-based velocity estimation for autonomous racing,'' in \emph{2021 IEEE
  International Conference on Autonomous Systems (ICAS)}, 2021, pp. 1--5.

\bibitem{chen2019kalman}
Z.~Chen, N.~Ahmed, S.~Julier, and C.~Heckman, ``Kalman filter tuning with
  bayesian optimization,'' \emph{arXiv preprint arXiv:1912.08601}, 2019.

\bibitem{shah2014student}
A.~Shah, A.~Wilson, and Z.~Ghahramani, ``Student-t processes as alternatives to
  gaussian processes,'' in \emph{Artificial intelligence and statistics}.\hskip
  1em plus 0.5em minus 0.4em\relax PMLR, 2014, pp. 877--885.

\bibitem{tracey2018upgrading}
B.~D. Tracey and D.~Wolpert, ``Upgrading from gaussian processes to
  student’st processes,'' in \emph{2018 AIAA Non-Deterministic Approaches
  Conference}, 2018, p. 1659.

\bibitem{torun2018bayesian}
H.~M. Torun, J.~A. Hejase, J.~Tang, W.~D. Beckert, and M.~Swaminathan,
  ``Bayesian active learning for uncertainty quantification of high speed
  channel signaling,'' in \emph{2018 IEEE 27th Conference on Electrical
  Performance of Electronic Packaging and Systems (EPEPS)}.\hskip 1em plus
  0.5em minus 0.4em\relax IEEE, 2018, pp. 311--313.

\bibitem{williams2006gaussian}
C.~K. Williams and C.~E. Rasmussen, \emph{Gaussian processes for machine
  learning}.\hskip 1em plus 0.5em minus 0.4em\relax MIT press Cambridge, MA,
  2006, vol.~2, no.~3.

\bibitem{clare2020expected}
C.~Clare, G.~Hawe, and S.~McClean, ``Expected regret minimization for bayesian
  optimization with student's-t processes,'' in \emph{Proceedings of the 2020
  3rd International Conference on Artificial Intelligence and Pattern
  Recognition}, 2020, pp. 8--12.

\end{thebibliography}


\end{document}